\documentclass[11pt,letterpaper]{article}
\usepackage{graphicx}
\usepackage[super,comma,sort&compress]{natbib}
\usepackage{amssymb}
\usepackage{geometry}
\usepackage{multicol}

\newif\ifusesup
\usesuptrue

\newcommand{\thetitle}{Cavity-aided magnetic-resonance microscopy of atomic transport in optical lattices}

\newcommand{\eqref}[1]{(\ref{#1})}
\newcommand{\dca}{\Delta_{ca}}
\newcommand{\dpa}{\Delta_{pa}}
\newcommand{\dpc}{\Delta_{pc}}
\newcommand{\sfrac}[2]{{^{#1}}\!/_{#2}}
\newcommand{\br}{\mathbf{r}}
\newcommand{\bs}{\mathbf{s}}

\newcommand{\dwrf}{\dot{\omega}_\mathrm{rf}}
\newcommand{\tmu}{{$\mu$}}
\newcommand{\micron}{$\mu$m}
\newcommand{\rb}{${^{87}}$Rb}
\newcommand{\dcaption}[1]{\caption{#1}}
\newcommand{\ket}[1]{\ensuremath{\left | #1 \right \rangle}}

\makeatletter
\renewcommand{\@makecaption}[2]{%
  \vskip\abovecaptionskip
  \sbox\@tempboxa{\textbf{#1: #2}}%
  \ifdim \wd\@tempboxa >\hsize
    \textbf{#1:} #2\par
  \else
    \global \@minipagefalse
    \hb@xt@\hsize{\hfil\box\@tempboxa\hfil}%
  \fi
  \vskip\belowcaptionskip}
\makeatother

\begin{document}

\twocolumn[
\title{\textbf{\textsf\thetitle}}
\author{
    \textsf{Nathan Brahms$^1$\thanks{Email: nbrahms@berkeley.edu}, Thomas P.\ Purdy$^{1}$\thanks{Present address:\ JILA, University of Colorado, Boulder, CO 80309, USA}, Daniel W.C.\ Brooks$^1$, Thierry Botter$^1$,}\\\textsf{and Dan M.\ Stamper-Kurn$^{1,2}$}
    \\ \\
    \small{1. Department of Physics, University of California, Berkeley, CA 94720, USA}\\
    \small{2. Materials Sciences Division, Lawrence Berkeley National Laboratory,}\\
    \small{Berkeley, CA 94720, USA}\\
}
\date{}

\maketitle
\thispagestyle{plain}
]{

\textbf{\textsf{
Ultracold atoms are emerging as an important platform for precision sensing and measurement, quantum information science, and simulations of condensed-matter phenomena.  Microscopic imaging is a powerful tool for measuring cold-atom systems, enabling the readout of ultracold atomic simulators \cite{bakr09microscope,Sherson2010} and registers \cite{schr04register}, the characterization of inhomogeneous environments \cite{Vengalattore2007:Magnetometry}, and the determination of spatially varying thermodynamic quantities \cite{shin08phasediagram,ho10phasediagram,navo10eqstate,hori10}.  Cold-atom microscopy has recently been demonstrated with imaging resolution sufficient to detect and address single \cite{kuhr:arxiv:addressing} or multiple \cite{ott:prl:addressing} atoms at individual optical-lattice sites with either micron- \cite{sche00co2lattice,nels07} or sub-micron-scale \cite{kars09nearest,bakr09microscope,Sherson2010,geri08ebeam} lattice spacing. However, such methods, which rely either on the fluorescence\cite{kuhr:arxiv:addressing,sche00co2lattice,nels07,kars09nearest,bakr09microscope,Sherson2010} or ionization\cite{geri08ebeam,ott:prl:addressing} of atoms, destroy the quantum states being measured and have limited dynamic range.  Here we demonstrate magnetic-resonance imaging of atomic gases in optical lattices, obtained by dispersively coupling atoms to a high-finesse optical cavity.  We achieve  state-sensitive, single-lattice-site images with high dynamic range.  We also apply this technique to measure the nonequilibrium transport dynamics of the gas.
}}

The sensitivity of optical cavities has been used to make nondestructive \cite{meschede:prl:qjs}, state-sensitive \cite{reichel:prl:cavbec}, and high-dynamic-range \cite{leroux:prl:squeeze} measurements of atomic gases, while magnetic resonance has been used to selectively address the spin states of single atoms in single optical-lattice sites \cite{kuhr:arxiv:addressing,meschede:njp:nmr}.  In this experiment we use light in a high-finesse optical cavity, together with radio-frequency (rf) radiation and an inhomogeneous magnetic field, to take real-time magnetic-resonance images (MRI) of atomic spins in an optical lattice.  We obtain a spatial resolution of 150 nm, far below the 425~nm spacing of the lattice sites.  We obtain a number-counting sensitivity of $\pm 10$ for up to 1000 atoms in each site or $\pm 2.5$ for up to 70 atoms, both well below the level of Poissonian atom-number fluctuations.  Furthermore, the MRI is minimally destructive of the atoms' internal states, allowing for the single-site observation of spatially dependent spin dynamics.  We use the technique to measure the transport dynamics of an initially localized gas via resonant quantum tunnelling between lattice sites, as the atoms undergo first ballistic and then interaction-inhibited transport.  We also demonstrate the ability to address the spins of large numbers of atoms at selected lattice sites, enabling new studies of magnetism, transport, cavity spin optodynamics \cite{brah10csod}, and cavity optomechanics \cite{murc08backaction,bren08opto,purd10tunable}.

\begin{figure}[t]
	\centering
	\includegraphics[width=\linewidth]{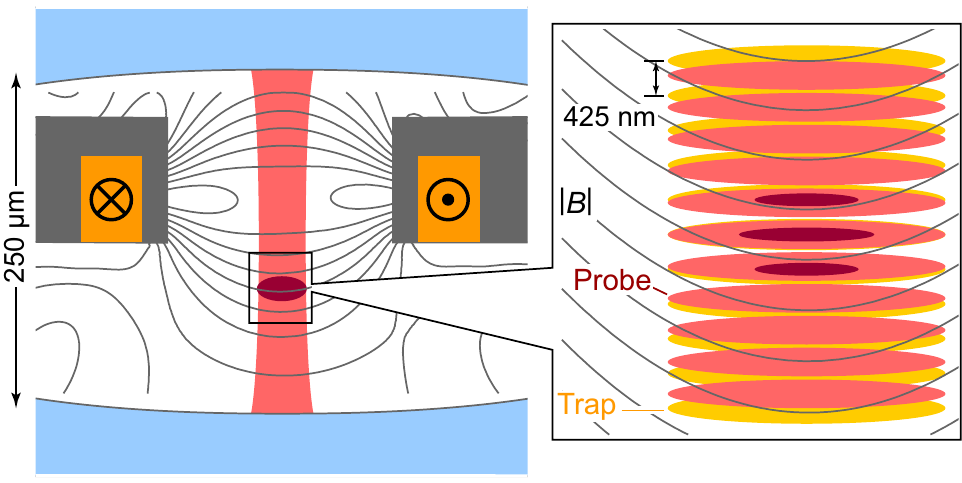}
	\dcaption{\textbf{An ensemble of ${^{87}}$Rb atoms optically trapped within a vertically oriented high-finesse Fabry-Perot cavity.}  Copper wires (orange, with current direction indicated) embedded within a 100-$\mu$m-thick silicon substrate (gray), together with an external bias coil, produce both a strong vertical magnetic-field gradient ($|B|$ contours shown) and a vertical bias field near the atoms. Inset: Atoms (red) are trapped at the antinodes of a standing-wave optical lattice (yellow) with 425~nm lattice spacing.  Circularly polarized cavity probe light (pink), detuned several gigahertz from the $\mathrm{D}_2$ line, acquires a dispersive phase shift that is sensitive to the atom spin projection along the cavity axis.}
	\label{fig:schematic}
\end{figure}

In contrast to traditional MRI, which measures transverse magnetisation at the Larmor precession frequency, here we use the cavity to perform a quantum-non-demolition measurement of the static longitudinal magnetisation \cite{kuzmich:pra:qnd} by applying a magnetic bias field along the cavity axis $\hat{k}$ (Fig.\ \ref{fig:schematic}).  To resolve the atoms spatially we apply a strong field gradient $B'$ along the dimension to be imaged.  For the data presented here we image along the cavity axis, although any axis can be imaged by applying the appropriate gradient.  We note especially that 3D images can be taken by imaging along a suitable number of axes and applying tomographic reconstruction \cite{rowland:tomography}.

For a sufficiently large detuning $\dca$ between the cavity resonance frequency and the atomic optical resonance, the absorption of cavity light by the atoms is negligible, and the light-atom interaction may be described by a real-valued index of refraction.  This index causes the cavity resonance frequency $\omega_c$ to be shifted from its empty-cavity value $\omega_0$.  Because the atoms are circularly birefringent, the resonance frequency of circularly polarized light is shifted by an amount $\Delta_N$, which depends on the atom density $\rho(\br)$ and the density $\bs(\br)$ of the dimensionless atomic spin (Supplementary Information):
\begin{equation}
\Delta_N \equiv \omega_c - \omega_0 = \int g(\br) \left [ \rho (\br) \pm C \bs(\br)\cdot \hat{k} \right ] d^3\br.
\label{eq:dntheory}
\end{equation}
Here $g(\br)$ is the scalar dispersive atom-cavity coupling parameter at position $\br$, taking into account the spatially varying intensity of the cavity mode.  The $\pm$ corresponds to $\sigma^\pm$ light polarizations, while $C$ is derived from a sum over probability amplitudes and relative detunings from the atomic excited hyperfine states.  For $^{87}$Rb atoms with hyperfine spin $F=2$ and probe light detuned by several gigahertz from the D$_2$ atomic resonance, $C \simeq \sfrac 14$. 

We invert the local magnetisation via adiabatic passage using a chirped rf field.  We use a linear chirp, with detuning from the magnetic resonance at position $z$ given at time $t$ by $\delta = \dwrf t - \mu B' z/\hbar$, where $\mu/\hbar$ is the atomic gyromagnetic ratio.  The radially integrated density $s_k(z)$ of the dimensionless atomic spin can be extracted from the time derivative of the cavity frequency.  For narrow magnetic resonance, where the spins at $z$ flip completely as the rf is swept from just below to just above resonance, the spin density is:
\begin{equation}
	s_k(z) = \mp\frac{1}{2 C \bar{g}(z)} \frac{\mu B'}{\hbar \dwrf}
		\left .\frac{d\omega_c}{dt} \right |_{t = \frac{\scriptstyle{\mu B' z}}{\scriptstyle{\hbar \dwrf}}}.
	\label{eq:skconv}
\end{equation}
Here $\bar{g}$ is the density-weighted radial average of the cavity coupling at $z$.  We neglect the small radial variation of the magnetic field across the gas.  This formula is easily refined to account for the finite frequency width of the magnetic resonance (Supplementary Information).

Experiments are conducted using a Fabry-Perot optical cavity integrated onto a microfabricated atom-chip device \cite{purd10tunable}.  The chip is used for sample preparation, delivering atomic gases of up to 5000 atoms, spin polarized in the $\ket{F,m_F} = \ket{2,2}$ hyperfine state, at temperatures of 1 to 3 $\mu$K into the optical lattice.  Chip wires are then used to apply strong magnetic field gradients, allowing us to resolve and address individual lattice sites.

The optical cavity is driven with two different wavelengths of light.  Light at a wavelength of 850 nm establishes a far-off-resonant optical lattice potential.  A cavity probe beam, detuned 14 to 17 GHz to the red of the D$_2$ atomic resonance at a wavelength of 780 nm, measures atom number and spin densities as described above.  The atoms are loaded into a few (between 2 and 5) adjacent lattice sites, centred on a site that overlaps with an antinode of the probe field.

\begin{figure*}[t]
	\centering
	\includegraphics{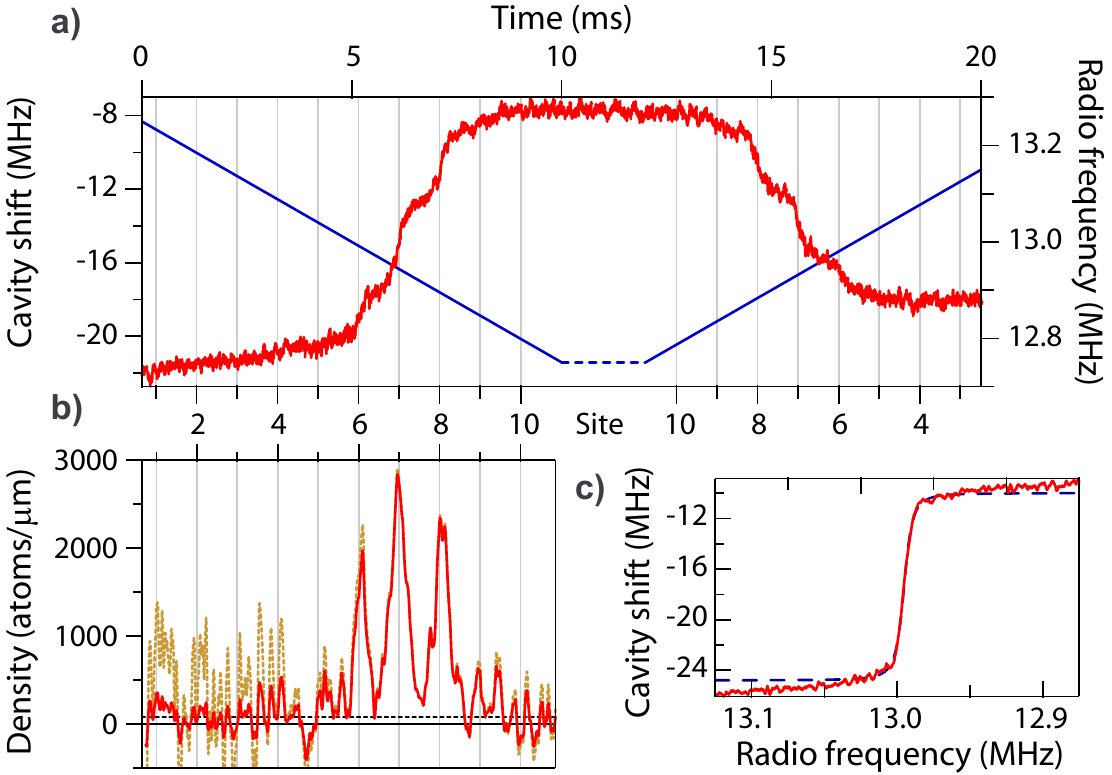}
	\dcaption{\textbf{Single-shot MRI of atoms in an optical lattice.} For this image, $\dca/2\pi = -14$~GHz with $1\times 10^7$~photons/s exiting the cavity, and 1800 atoms initially in the $\ket{F,m_F} = \ket{2,2}$ state.  (A) The shift $\Delta_N \equiv \omega_c-\omega_0$ of the cavity from its empty resonance frequency (red line, left axis), showing steps as the rf (blue line, right axis) is chirped from high to low.  As the rf is swept back, the detuning recovers its initial value with 85\% each-way fidelity. (B) Atomic density as calculated from the time derivative of $\Delta_N$, both uncorrected (red solid line) and corrected (yellow dotted line) for spatially varying sensitivity (probe and trap antinodes are overlapped at site 7).  The peak widths (200~nm FWHM) are given by the convolution of the imaging resolution (150~nm), the atom-distribution width (100~nm), and a low-pass analysis filter (90~nm).  The MRI has an 80 atom/$\mu$m offset due to deterministic atom loss (dotted line).  (C)  The imaging resolution is given by the ratio of the 14~kHz magnetic-resonance width to the 114~kHz/$\mu$m magnetic-field gradient.  The magnetic-resonance width is measured by sweeping over resonance in a uniform bias field (red solid line) and fitting to adiabatic-passage theory (blue dashed line).}
	\label{fig:mrispatial}
\end{figure*}

To measure the cavity shift we detect the photon flux of $\sigma^+$-polarized probe light transmitted through the cavity.  A feedback loop tunes the probe frequency $\omega_p$, maintaining the flux at a constant value $\bar{\gamma}$ equal to a fixed fraction of the incident photon flux. The probe frequency is thus locked at a fixed detuning $\dpc$ from cavity resonance, the frequency of which is determined by $\omega_c = \omega_p - \dpc$. Variations in the cavity resonance frequency faster than the 20~kHz feedback bandwidth can be assessed by using the residual deviation of the instantaneous photon flux $\gamma$ from $\bar{\gamma}$.

Fig.~\ref{fig:mrispatial} shows a typical single-shot image of a spin-polarized atomic gas taken with a linear rf chirp. As the rf is swept through the resonance of each lattice site, atoms are flipped from the $\ket{2,2}$ to the $\ket{2, -2}$ state, causing a jump in the cavity resonance frequency. The atom density is obtained by calculating $s_k(z)/2$ using Eq.~(\ref{eq:skconv}).  We account for the thermal radial and axial distribution of atoms within each lattice site in determining $\bar{g}(z)$.  We also account for the difference between the trap and probe light wavelengths, although this latter correction is minor for lattice sites near the common antinode of the two cavity modes.  MRIs may also be averaged (see Fig.~\ref{fig:diffmri}), although shot-to-shot variations in the bias field cause broadening of the averaged images.  The flatness of the slope of the cavity shift between spin flips verifies our ability to address single lattice sites (see Supplementary Information).

The spatial resolution with which we can image the spins is 150~nm (full width at half-maximum).  The resolution is proportional to the ratio of the magnetic-resonance width (14~kHz, see Fig.~\ref{fig:mrispatial}c) to the field gradient (114~kHz/$\mu$m).  The minimum resolution is limited by two considerations.  First, the Rabi frequency must be sufficiently high that the spins are inverted adiabatically.  Secondly, the maximum magnetic-field gradient is limited by transverse field curvatures, which expel atoms from the optical trap (see Supplementary Information).

\begin{figure}[t]
	\centering
	\includegraphics[width=\linewidth]{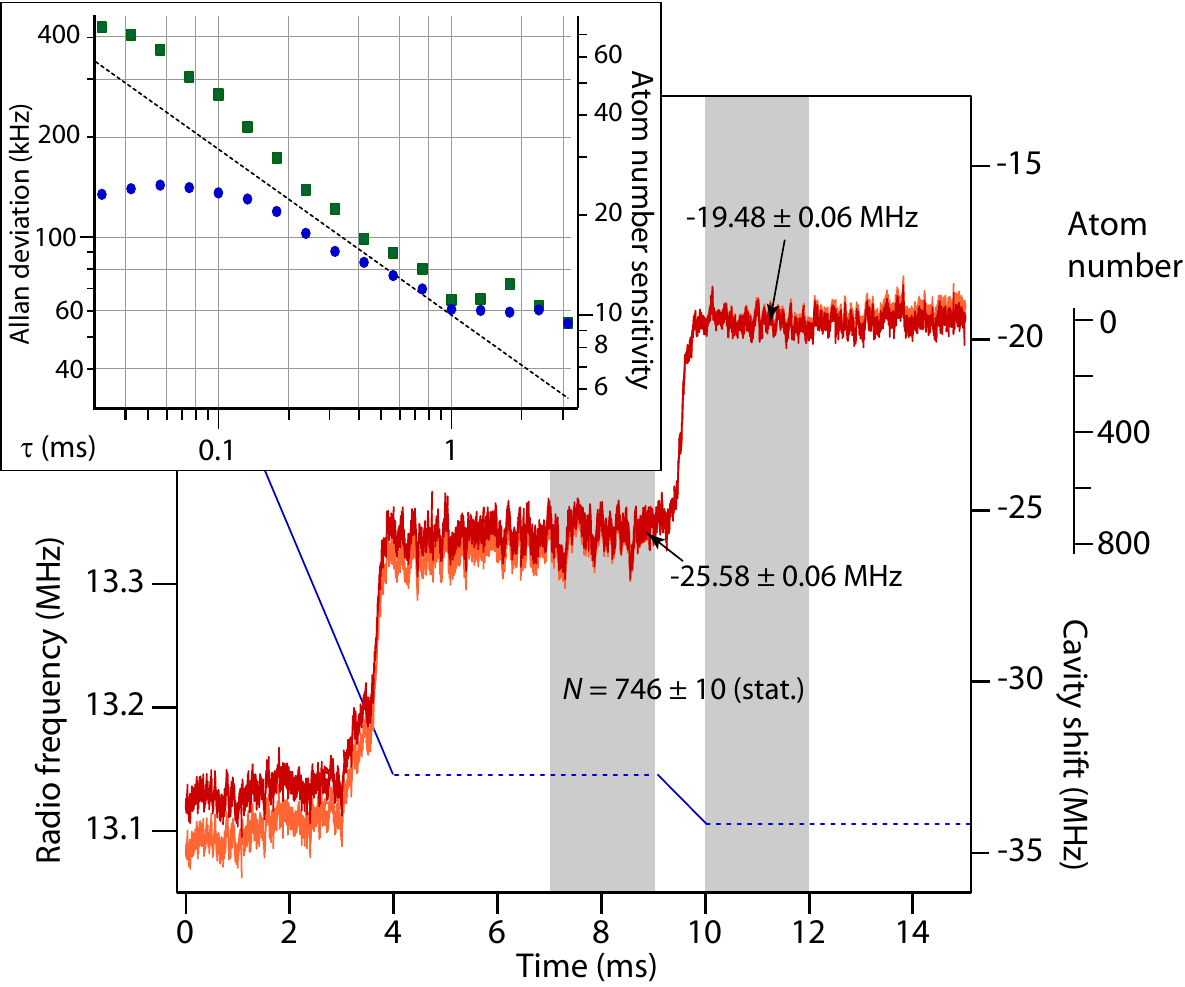}
	\dcaption{\textbf{Single-shot measurement of atom number in a single lattice site.} The rf sweep (blue, left axis, solid line at full amplitude, dashed line at zero amplitude) is halted for a measurement period (shaded area), then swept across the lattice site of interest, before halting and measuring again.  The atom number is calculated from the cavity shift (orange, right axis) after correcting for deterministic atom loss (red, right axis, corrected about $t=9.5$~ms).  Here $\dca/2\pi = -14$~GHz, corresponding to 122 atoms/MHz of cavity shift.  The inset shows the Allan deviation of the cavity shift after correcting for atom loss, using the probe frequency only (blue circles) and corrected using the instantaneous cavity transmission (green squares).  The dotted line is the expected deviation due to photon shot noise.  The deviation on the probe frequency measurement at short times is below the shot-noise expectation due to a 20~kHz electronic filter.  The 60~kHz Allan deviation for $\tau=2.2$~ms corresponds to a sensitivity of 10 atoms.}
	\label{fig:singlewell}
\end{figure}

The atomic magnetisation is largely preserved during the imaging process and can be recovered by reversing the rf chirp, as shown in Fig.~\ref{fig:mrispatial}.  The magnetisation is slightly diminished due to radiation pressure fluctuations caused by the probe light \cite{murc08backaction}, which reduce the trapping lifetime to 120 ms, and due to spin decoherence and gradient-induced loss associated with executing the spin flips.  Together, these processes result in $85\%$ of the magnetisation being preserved during the image.  Less destructive imaging could be achieved at the expense of reduced signal-to-noise ratio or spatial resolution, e.g.\ by decreasing $\bar{\gamma}$, increasing the rf drive strength, or decreasing $B'$.

Precise measurements of the longitudinal spin in a single lattice site can be taken by measuring the cavity resonance frequency before and after the spins in a single site are inverted (Fig.\ \ref{fig:singlewell}).  The resonance frequency is first measured for a time $\tau$ with the rf off.  The rf is then adiabatically turned on  (125~\tmu s linear ramp-on) at a frequency 20~kHz above the spin-resonance frequency, chirped to 20~kHz below the spin-resonance frequency, and finally adiabatically ramped off.  Finally, $\Delta_N$ is again measured, and the total projection $S_k$ of the site's dimensionless spin is calculated using $S_k = \Delta \omega_c/2 C \bar{g}$, where $\Delta\omega_c$ is the change in the cavity resonance frequency.

For small $\tau$, the precision of the single-site measurement is limited by the photon shot noise on the cavity resonance measurement to
$\delta \omega_c = |d\omega_c/d \gamma| \times \sqrt{\bar{\gamma}/\varepsilon \tau}$,
where $\varepsilon$ is the photodetection quantum efficiency and  $|d\omega_c/d\gamma|$ is the resonance frequency measurement sensitivity, $\simeq 11$ for our system (see Supplementary Information).  For $\tau$ above 1 ms, a measurement of the Allan deviation (Fig.~\ref{fig:singlewell}) of $\omega_c$ indicates that photon shot noise is superseded by technical noise, e.g.\ variations of the laser frequency or of $\omega_0$, yielding a frequency uncertainty of 60 kHz for $\tau = 2.2$ ms.  With $\dca=-14$~GHz, the atom number in a single lattice site is thus determined with an rms uncertainty $\delta N = 10$. This sensitivity should be sufficient to observe atom-number differences between lattice sites \cite{este08squeeze} below the limit of Poissonian statistics ($\delta N \sim 30$ for $N \sim 1000$), while maintaining the probed gas for further experiments.  We have also measured populations to $\delta N = 2.5$, by using $\dca=-2.0$~GHz, although at this detuning the dynamic range of our measurement is limited to $N\lesssim 70$ within each site.  Systematic uncertainties (Supplementary Information) affect the accuracy of the single-site measurement on the few-percent level but do not significantly impact the measurement precision.

\begin{figure*}[t]
	\centering
	\includegraphics[scale=0.9]{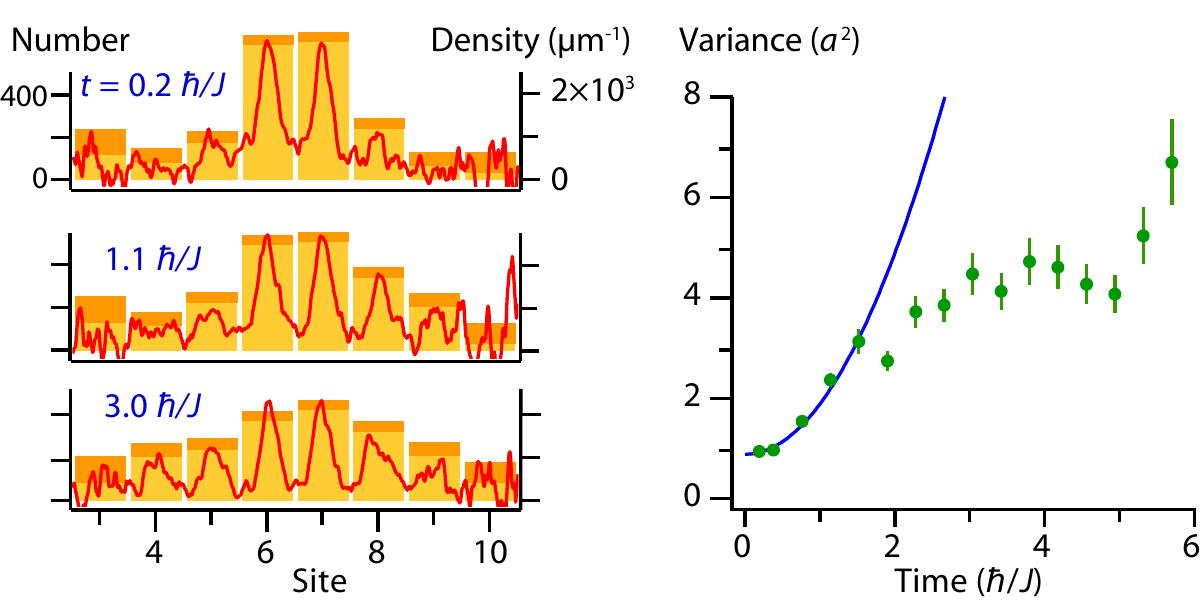}
	\dcaption{\textbf{Atom transport in an optical lattice as measured using cavity-aided MRI.}  ${^{87}}$Rb atoms at 380~nK tunnel resonantly in a $10.1~E_r$ lattice with tunnelling matrix element $J=\hbar\times380~\mathrm{s}^{-1}$.  Left:  After allowing the atoms to evolve for a fixed time ($t = 0.2,~1.1,~3.0~\hbar/J$ shown), we take an MRI, corrected for spatially varying sensitivity.  For each evolution time, 15 MRIs are averaged together (red line, right axis), and these are integrated to obtain the atom number distribution among lattice sites (yellow bars, left axis, orange region indicates 68\% certainty as obtained from Allan deviation).  Each distribution is then fit to a gaussian envelope.  Right:  Position variance $\sigma^2$ of the gaussian envelope fit, expressed in units of square lattice spacing, as a function of transport time, expressed in units of inverse tunnelling rate.  At early times ($t<2\hbar/ J$), the data (green circles, error bars denote 68\% certainty from fits) agree with no-free-parameter ballistic tunnelling theory (blue line).}
	\label{fig:diffmri}
\end{figure*}

We use the MRI to probe the transport dynamics of an atomic gas in an optical lattice under the effects of atomic interactions.  Here we begin with atoms localized to a few lattice sites and use the MRI to image the initial stages of their expansion.  We prepare a nondegenerate spin-polarized sample at 380~nK, with an initial width of 380~nm.  We allow the gas to evolve for variable time $t$ in a shallow optical lattice, which has a potential depth of $10.1~E_r$, where $E_r = h \times 3.17$~kHz is the rubidium recoil energy at a wavelength of 850 nm.  The lattice depth is then raised to take the MRI.  The gravitational force is compensated by applying a weak levitating magnetic field gradient ($\mu B'/h = 454~$Hz per lattice constant) during transport.  The experiment is repeated 15 times at each of several different values of $t$, and the MRIs for each hold time are averaged together, after correcting for shot-to-shot variations of the bias field (Fig.~\ref{fig:diffmri}).

The position variance of the atomic distribution grows due to quantum tunnelling.  At early times ($t\leq 4$~ms) the growth matches the ballistic expansion of non-interacting atoms in the lattice, for an initial atomic axial distribution given by an incoherent mixture of single-site Wannier states.  At later times the expansion slows dramatically.  Previous experiments have observed self trapping in Bose-Einstein condensates \cite{oberthaler:prl:selftrap}.  The behavior we observe agrees qualitatively with our simulations of interaction-induced self-trapping in a nondegenerate gas, enabled in this experiment by high atom density.

Using dispersive optical measurement, cavity enhancement, and magnetic resonance, we have demonstrated a spin-sensitive technique for imaging atomic gases with a spatial resolution of 150~nm.  The single-shot atom number sensitivity is as small as 2.5 within a single lattice site, low enough to observe sub-Poissonian statistics of atom-number differences between sites with more than 10 atoms.  Extending this technique to single-spin and single-atom counting should be possible by reducing technical measurement noise and increasing the experimental photon detection efficiency, currently limited by cavity losses.

The ability to control and to read out spins in single lattice sites provides new tools to engineer and study atomic gases at a microscopic level.  Here we have measured the initial dynamics of bosonic atom transport in a 1D lattice with single-site resolution.  Other experiments could include the observation of cavity-mediated long-range interactions between independently prepared spin populations or the observation of individual magnetic domains for antiferromagnetically ordered systems.  Finally, because our dispersive measurement can be made minimally destructive, the method could be used to provide realtime feedback to spin projections of individual lattice sites for studies of quantum measurement and control \cite{Jessen2009:PRL:MicrowaveControl}, and for applications to atomic magnetometry.

{
\small
\setlength{\bibsep}{2pt}

{\raggedright 
\paragraph*{Acknowlegements} This work was supported by NSF Grant \#PHY-0801827 and the AFOSR.  D.M.S.-K.\ acknowledges support from the Miller Institute for Basic Science, and T.B.\ acknowledges support from Le Fonds Qu\'eb\'ecois de la Recherche sur la Nature et les Technologies.  The authors thank A.M.\ Rey and K.\ Hazzard for discussions on the transport experiment.
\paragraph*{Author Contributions} Experimental data were taken by T.P., N.B., D.W.C.B., and T.B.  All authors were involved with experimental design, data analysis, and production of the manuscript.
}

}
}

\clearpage

\setlength\oddsidemargin{0.50in}
\setlength\topmargin{0.0in}
\setlength\textheight{8.0in}
\setlength\footskip{0.75in}
\onecolumn{
\setlength\hsize{5.5in}
\setlength\columnwidth{\hsize}
\setlength\linewidth{\hsize}
\ifusesup
	\setlength\baselineskip{16pt}

\section*{Supplementary Information}

\setcounter{equation}{0}
\setcounter{figure}{0}
\setcounter{page}{1}
\renewcommand{\thepage}{\hspace{-1.75in}\roman{page}}
\renewcommand{\bibnumfmt}[1]{[S#1]}
\newcommand{\citesup}[1]{\bibpunct{S}{}{,S}{s}{}{}\cite{#1}\bibpunct{}{}{,}{s}{}{}}

\paragraph*{Atom cooling and optics.} The experimental procedure is similar to that of Ref.~\citenum{purd10tunable}.  We begin with a magneto-optical trap of \rb, loaded from a background vapor.  After polarization-gradient cooling and optical pumping to the $\ket{F,m_F} = \ket{2,2}$ state, $10^7$ atoms are loaded at $\approx 20~\mu\textrm{K}$ into a magnetic trap created using wires on the atom chip.  Evaporative cooling and magnetic transport is then used to create a $\mathrm{4\times0.4\times0.4~\mu m}$ cloud, located within the Fabry-Perot optical cavity.  The optical lattice is then ramped on and the magnetic trap ramped off.  After a final stage of evaporative cooling within the optical lattice, we end with 1500 to 5000 atoms at 1 to 3 $\mu$K, in two to five lattice sites of a $410~E_r$-deep lattice.  The cavity mirrors (Research Electro-Optics, Inc.) are spaced by 250~\micron\ and have 5~cm radii of curvature, giving $g_\mathrm{max}/2\pi = 11$~kHz when $\dca/2\pi = 10$~GHz.  The mirrors have a finesse of 160,000 at 780 nm, yielding resonances with half-linewidth $\kappa/2\pi = 1.8$~MHz.  Both the trap and probe lasers drive TEM$_{00}$ modes of the cavity.  Each of these spatially gaussian modes has a minimum $1/e^2$ diameter of $\sim 50~\mu\mathrm{m}$.  The probe laser is locked in tandem with the lattice laser to a separate, thermally and mechanically stabilized confocal Fabry-P\'erot transfer cavity, with 30~kHz relative RMS noise in a 2~MHz bandwidth.

The probe intensity is stabilized to a constant value before entering the science cavity.  To measure the science cavity resonance, the photon flux leaving the cavity is stabilized to $\bar{\gamma}$ by feedback to the probe frequency via an acousto-optic modulator, maintaining a constant detuning $\dpc$ from cavity resonance.  For typical operating parameters, $\bar{\gamma} \sim 10^6~\mathrm{s}^{-1}$.  The feedback loop is integral, with a 8~$\mu$s time constant.  The instantaneous value of the cavity resonance frequency is calculated using $\omega_c = \omega_\mathrm{XC} + \omega_\mathrm{AOM} - \dpc + (\bar{\gamma} - \gamma) (\dpc^2 + \kappa^2) / 2 \dpc \bar{\gamma}$. Here $\omega_\mathrm{XC}$ is the frequency of the probe laser locked to the transfer cavity and $\omega_\mathrm{AOM}$ is the frequency feeding the acousto-optic modulator.  $\dpc/2\pi$ is typically chosen to be $-2.1~\mathrm{MHz}$.  The sensitivity of the cavity resonance measurement to the transmitted photon flux is determined from the Lorentzian cavity transmission lineshape, and is $d\omega_c/d\gamma = (\dpc^2 + \kappa^2) / 2 \dpc \gamma $. 

\paragraph*{Cavity-aided spin measurement.} For circularly polarized light, with detuning from atomic resonance $\dpa$ much greater than the atomic natural linewidth, and with all atoms in the same hyperfine manifold, the value of the atomic index of refraction $n_r$ depends on the densities $\rho_m(\br)$ of atoms in each magnetic sublevel $m$ as \citesup{CCT,happer:prl:offresonant}:
\begin{equation}
n_r \simeq 1 - \frac{|d_{12}|^2}{2 \hbar \varepsilon_0} \sum_{m, F'} \rho_m(\br) \frac{|V(F,m|F',m\mathord{\pm}1)|^2}{\hbar \omega_p - E_{F'} + E_F}.
\label{eq:nrtheory}
\end{equation}
Here $d_{12}$ is the atomic transition dipole matrix element and the $\pm$ corresponds to $\sigma^\pm$ light.  $V(F,m|F',m\mathord{\pm}1)$ describes the transition probability amplitude from $\ket{F,m}\rightarrow\ket{F',m \pm 1}$ using $\sigma^\pm$ light, and $E_{F'}$ and $E_F$ are the energies of the excited and ground hyperfine states, respectively.  If $\dpa$ is large compared to the excited-state hyperfine splitting, we can replace $\hbar \omega_p - E_{F'} + E_F \simeq \dpa$.  The sum over hyperfine states in Supplementary Eqn.~\eqref{eq:nrtheory} now becomes $(\xi+\upsilon m)/\dpa$, i.e.\ a scalar and a vector dependence on the spin density, with negligible tensor contribution. For $F=2$ rubidium-87, $\xi=\sfrac 23$ and $\upsilon=\sfrac 16$. 

The resonance frequency of a cavity mode, of volume $V_0$, is shifted by an intracavity refractive medium by
\begin{equation}
\omega_c = \frac{\omega_0}{\bar{I} V_0} \int \! \frac {I(\br)}{ n_r(\br)} \, d^3\mathbf{r} ,
\label{eq:wctheory}
\end{equation}
where $I(\br)$ is the intensity of the cavity mode at position $\br$ with average value $\bar{I}$ and the integral is evaluated with $\dpa = \dca$.  For $n_r$ near unity, Supplementary Eqn.~(\ref{eq:wctheory}) becomes Article Eqn.~(\ref{eq:dntheory}), writing the dispersive cavity-coupling parameter 
\begin{equation}
g(\br) \equiv \frac{|d_{12}|^2 \xi \omega_0 I(\br)}{2 \hbar \varepsilon_0 \dca  V_0 \bar{I}},
\end{equation}
together with $C \equiv \upsilon/\xi$.

To calculate the radially averaged coupling parameter $\bar{g}$, we begin with the spatial dependence of
\begin{displaymath}
	g(\br) = g_\mathrm{max} \exp(-r^2/r_0^2) \cos(k_p (z-z_0))^2,
\end{displaymath}
where $g_\mathrm{max}$ is the value of $g(\br)$ at a location of maximum probe intensity, and $r_0$ is the $1/e^2$ radius of the probe intensity.  We now average this coupling over the spin-density distribution.  We assume that the spin density in a lattice site is proportional to the atomic density in that site, and write the atomic density using Boltzmann statistics for the transverse dimension $r$, giving $\rho(r,z) = \rho_z(z) \times  \exp(-U_r(r)/k_B T)$, where $U_r = U_0 (1-\exp(-r^2/r_0^2))$ is the transverse trap potential.  The atoms are cold enough that we can approximate $U_r/U_0 \approx r^2/r_0^2$. We now calculate $\bar{g}$ using the atom-density-weighted transverse average of $g(\br)$, giving
\begin{displaymath}
	\bar{g} = \frac{\int r g(\br) n_a(r,z) dr}{\int r n_a(r,z) dr} =
		g_\mathrm{max} \frac{U_0}{k_B T + U_0} \cos^2(k_p (z-z_0)),
\end{displaymath}
where $k_B T$ is the thermal energy of the transverse degrees of freedom.

The wavelength difference between the trap and probe fields causes a spatially modulated overlap between the atoms, trapped at the antinodes of the lattice, and the antinodes of the probe field, with an overlap period of 11 lattice sites\cite{purd10tunable}.  To avoid amplifying noise at the probe nodes, we can further average the correction over the axial distribution in a single lattice site.  For cold atoms in a tight confining potential, the distribution can be calculated using Bose-Einstein statistics for an harmonic oscillator, giving $\rho_z(z) = \rho_0 \exp[-m\omega_z z^2/(n_\mathrm{th}+\sfrac 12)\hbar]$, where $n_\mathrm{th} = [\exp(\hbar \omega_z/k_B T) -1 ]^{-1}$ is the thermal occupation of the axial mode.  Averaging over the distribution in a single site gives
\begin{equation}
	\frac{\bar{g}}{g_\mathrm{max}} = \frac{1}{2}\frac{U_0}{k_B T\! + U_0}
		\left [ 1 + e^{-\frac{\left ( n_\mathrm{th}+ \sfrac 12 \right )
			\hbar \omega_z}{ U_0 (k_t/k_p)^2}} \cos (2 k_d (z\mathord{-}z_0) )
		\right ]\!.
	\label{eq:gbar}
\end{equation}
Here $\omega_z$ is the axial frequency in each site, $n_\mathrm{th}$ is the thermal occupation of the axial mode, $k_t$ and $k_p$ are the trap and probe wavenumbers, $k_d = k_t-k_p$, and $z_0$ is the location of the lattice site which overlaps with the probe antinode.  For typical experimental parameters, $U_0/h = 1.28$~MHz, $\omega_z/2\pi = 135~\mathrm{kHz}$, and $U_0/k_B T \approx 20$.  The atoms begin with $n_\mathrm{th} = 0.15$.  This phonon number grows due to optomechanical heating of the atoms to approximately 0.5 (as measured from the MRI peak width -- although this calculation does not take into account motional narrowing of the MRI peaks).  These parameters give a $\bar{g}(z)$ that oscillates sinusoidally from approximately 5\% to 95\% between positions where the optical lattice site is overlapped with a node or an antinode of the probe field.  Note that the technique is thus sensitive to atoms at all lattice sites, although atoms at the node are measured with a signal-to-noise that is 20 times smaller than those at the antinode.

\paragraph{Spin addressing.} Under adiabatic-passage theory \citesup{rabi:rmp:rwa}, the expectation value of the spin-projection operator $\hat{S}_z$ of a two-level spin system, with energy splitting
$\hbar\Omega$ and detuning $\delta$ of the rf from magnetic
resonance, is
\begin{equation}
    \langle \hat{S}_z(\delta) \rangle = \frac{\delta}{\sqrt{\delta^2 + \Omega^2/4}} \langle \hat{S}_z \rangle_0,
    \label{eq:aptheory}
\end{equation}
where $\langle \hat{S}_z\rangle_0$ is the spin projection when $\delta/\Omega \rightarrow \infty$. As the rf is chirped, with $\delta = \dwrf t - \mu B(z)/\hbar$, the time-derivative of the cavity resonance obeys:
\begin{equation}
\frac{d \omega_c}{dt} = \frac{d\Delta_N}{dt} = \pm C \int \!\frac{\dwrf \Omega^2/4}{\left[\delta(t,z)^2 + \Omega^2/4\right ]^{3/2}} s_k(z) \bar{g}(z) dz,
\label{eq:dwc}
\end{equation}
where $s_k(z)$ is the initial one-dimensional spin density (at $t=-\infty$).  In the limit that $\hbar \Omega$ is small compared to $\mu B' s_k / (ds_k/dz)$, the detuning-dependent term in Supplementary Eqn.~\eqref{eq:dwc} becomes a delta function with magnitude $2 \hbar \dwrf/\mu B'$.  Evaluating the integral with the delta function gives the result of Article Eqn.~\eqref{eq:skconv}.  When $\Omega$ is not small, the integral in Supplementary Eqn.~\eqref{eq:dwc} acts as a point-spread function for the image, with full-width at half-maximum $= 1.23\; \hbar\Omega/\mu B'$.  We extract $\Omega$ in our system by fitting a sweep over magnetic resonance, at constant bias field, to Supplementary Eqn.~(\ref{eq:aptheory}), as shown in Fig.~{\ref{fig:mrispatial}}(c).

\begin{figure}[t]
	\centering
	\includegraphics{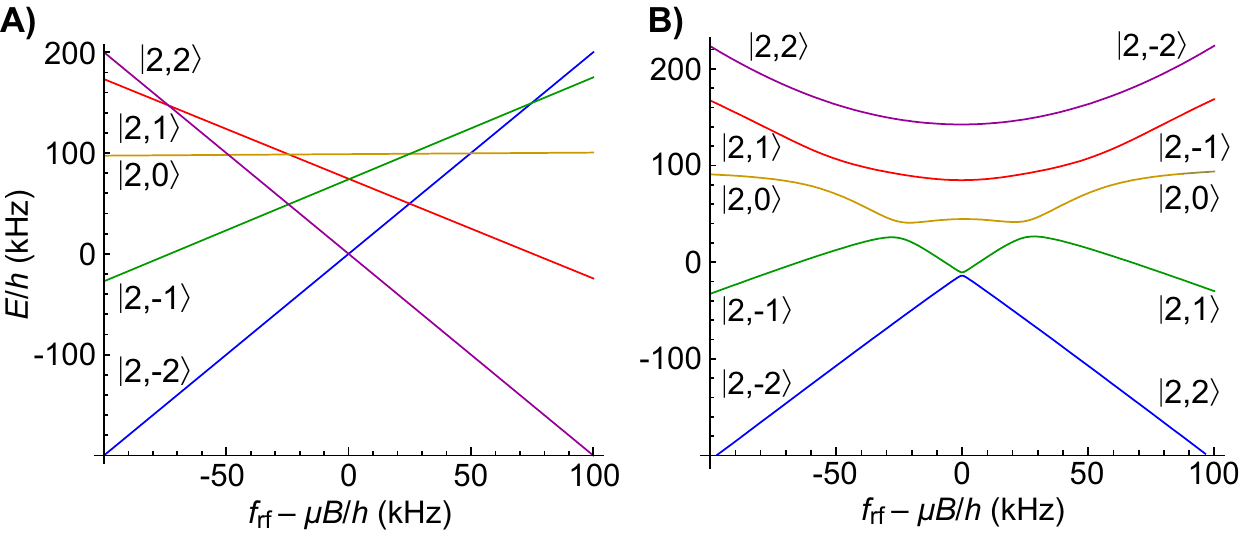}
	\dcaption{Calculations of the dressed Zeeman spectrum of $F\mathord{=}2$ \rb\ about rf resonance, as a function of Larmor frequency.  Calculations are for a bias field of $\mu B/h = 13~\textrm{MHz}$.  (a) Dressing with $\mu B_\bot/h = 0$~Hz.  An rf sweep from low to high frequency takes a gas starting in $\ket{F,m_F} = \ket{2,2}$ to $\ket{2,-2}$ along the upper set of crossings, with a crossing every 50~kHz.  Sweeps from high to low frequency take the $\ket{2,2}$ state to the $\ket{2,-2}$ state through only one avoided crossing.  (b) Dressing with $\mu B_\bot/h = 120$~kHz.  Crossings between states of $\Delta m_F = 1$ are split by $\mu B_\bot/2$, while the lower crossing between $\ket{2,2}$ and $\ket{2,-2}$ is split by only a small energy difference.}
	\label{fig:dressed}
\end{figure}

Due to the quadratic Zeeman shift \citesup{Breit31:QuadraticZeeman} present in $F=2$ \rb, the energy splitting $\Omega$ depends on the initial state of the atoms and the direction of the rf sweep. Beginning in the $\ket{F,m_F} = \ket{2,2}$ state, and chirping the rf from low to high frequency, the spin encounters a series of four broad resonances (cf.\ Supplementary Fig.~\ref{fig:dressed}), where states of $\Delta m_F =1$ cross, each of width $\mu B_\bot/2$, where $B_\bot$ is the effective transverse field in the rotating wave approximation.  These resonances are separated by twice the quadratic Zeeman shift, equal to 25~kHz at our operating bias field of 13~MHz.  In order to avoid these resonances, we instead chirp the rf from high to low frequency.  In this mode, only one crossing exists, between the initial $\ket{2,2}$ and the final $\ket{2,-2}$ state.  Because the connection between these states relies on a four-photon transition, it is much narrower, with $\Omega/2\pi = 14$~kHz achieved using $\mu B_\bot/ h = 200$~kHz at a 13~MHz bias field.

To produce $B'$, we run current in opposite directions along two parallel chip wires, which are oriented transverse to the cavity axis and each separated from the atoms by approximately 160~\micron\ (see Fig.~\ref{fig:schematic}).  With $\approx 1$~A in each wire, a field, oriented along the cavity axis, is created with axial gradient $\mu B'/h = 113.5$~kHz/\micron.  By adding a bias field along the cavity axis, we tune the mean field seen by the atoms between 10 and 13~MHz.  The rf for adiabatic passage is added through a bias tee on one of the chip wires.

\paragraph*{Single-site addressability.}
We are able to manipulate spins at a single lattice site, with a $\lesssim 6\%$ contamination of the magnetization at adjacent lattice sites.  We demonstrate this by measuring the time derivative of the cavity shift as the rf is swept between adjacent lattice sites, after correcting for atom loss.  The time derivative is a direct measure of the rate of change in magnetization per unit time (cf.~Eqn.~(\ref{eq:skconv})), and a small derivative indicates that few spins are being perturbed when the rf is between lattice sites.

Supplementary Fig.~\ref{fig:sup:address} shows our measurement of addressability.  Comparing the ratio of the between-site time derivative to the on-site time derivative, we obtain an upper limit for the amount by which spins at neighboring lattice sites are affected by the rf.  Measuring before the rf is swept over the most-occupied site, we measure a slope that is $2\%$ of the peak (average over 10 shots, statistical uncertainty is $\pm 2\%$).  Measuring after the sweep over this lattice site, the slope is $6\%$, but this increase can be attributed to decoherence of spins which were incompletely inverted during the sweep.

Note that this measurement gives an upper limit on the addressability, and the contamination is likely less than $6\%$.  Because the change in magnetization of neighboring lattice sites is due primarily to dressing of the spins by the rf, adiabatically turning the rf on or off should reverse much of the change in spin.

Due to fluctuations in the magnetic bias field (on the order of 6~kHz), the deterministic addressibility may be degraded.  This effect can be compensated for either by post-selection or by first measuring the bias field with an rf sweep.

\begin{figure}[t]
	\centering
	\includegraphics{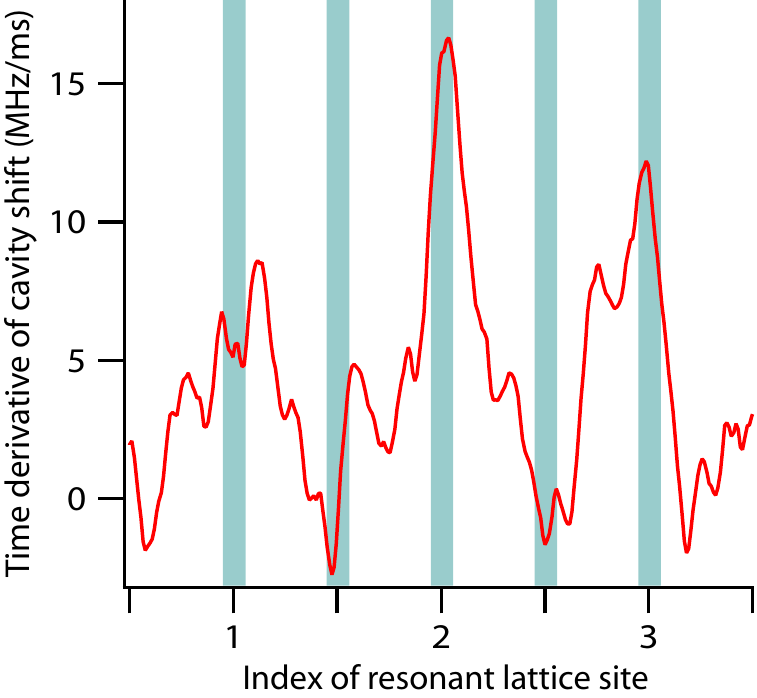}
	\caption{Measurement of single-site addressability.  The rate of spin inversion is proportional to the time derivative of the cavity shift (single-shot time derivative shown in red solid line).  An upper limit on the extent to which spins in neighboring sites are perturbed by addressing a single site is given by the ratio of the time-derivative between sites to the time-derivative on resonance.  The times we used to measure the derivative both on and off resonance are denoted by the shaded regions.}
	\label{fig:sup:address}
\end{figure}

\paragraph*{Spatial resolution.} The spatial resolution of the MRI is limited by the strength of the rf field used to invert the atomic spins and by the magnetic field gradient $B^\prime$ used to differentiate neighboring lattice sites.  Letting the energy $\hbar \Omega$ characterize the rf-induced level splitting between opposite spin projections, the spatial resolution (full-width at half-maximum) is limited to $1.23\; \hbar \Omega / \mu B'$.  The minimum value of the coupling rate $\Omega$ is limited by the requirement for adiabatic passage between opposite spin states ($\Omega^2 \gg |\dwrf|$) and by the requirement that the rf be swept across resonance in a time short compared to the transverse-spin-coherence time, measured to be $2$ ms for our system.  In practice, we find that using $\Omega/2\pi=14$~kHz and $\dwrf/2\pi = - 50$~kHz/ms maintains a 95\% spin coherence during the sweep. Meanwhile, the maximum value of $B'$ is around 1.7 kG/cm, separating neighboring sites by 50~kHz.  This value is limited by the transverse magnetic-field curvature, having value $\sim B'^2/B$, associated with applying the gradient.  If the gradient is too large, this curvature will eject $m_F\mathord{=}-2$ atoms from the resulting hybrid optical-magnetic trap.

\paragraph*{Atom counting.} Systematic uncertainties limit the accuracy (but not the precision) of single-site atom number measurements.  These uncertainties arise from uncertainty in $\dca$ (measured to within 50~MHz), variation in $\bar{g}$ due to temperature uncertainty (measured by time-of-flight methods to within $10\%$), and impure initial spin preparation (purity uncertainty $1\%$) due to either an atomic spin mixture or a bias field not exactly parallel to the cavity axis.

The numbers measured via the MRI technique were verified using two independent methods.  First, we have used time-of-flight absorption imaging to verify the total atom number in the optical lattice.  Second, we have compared the total atom number by using the scalar cavity shift from its empty resonance.  This latter test is equivalent to verifying that the shift with all spins in the $|2,2\rangle$ state is $\left ( \frac{1+2 C}{1-2 C} \right ) = 3$ times larger  than with all spins flipped to their $|2,-2\rangle$ state.

\paragraph*{Number sensitivity.}  The number-counting sensitivity is technically limited by fluctuations in the measured empty cavity resonance frequency, detection efficiency, and atom loss.  The dominant of these three effects is technical variation in the measured empty cavity frequency, due to acoustic vibrations of the cavity mirrors. 

\begin{figure}[t]
	\centering
		\includegraphics{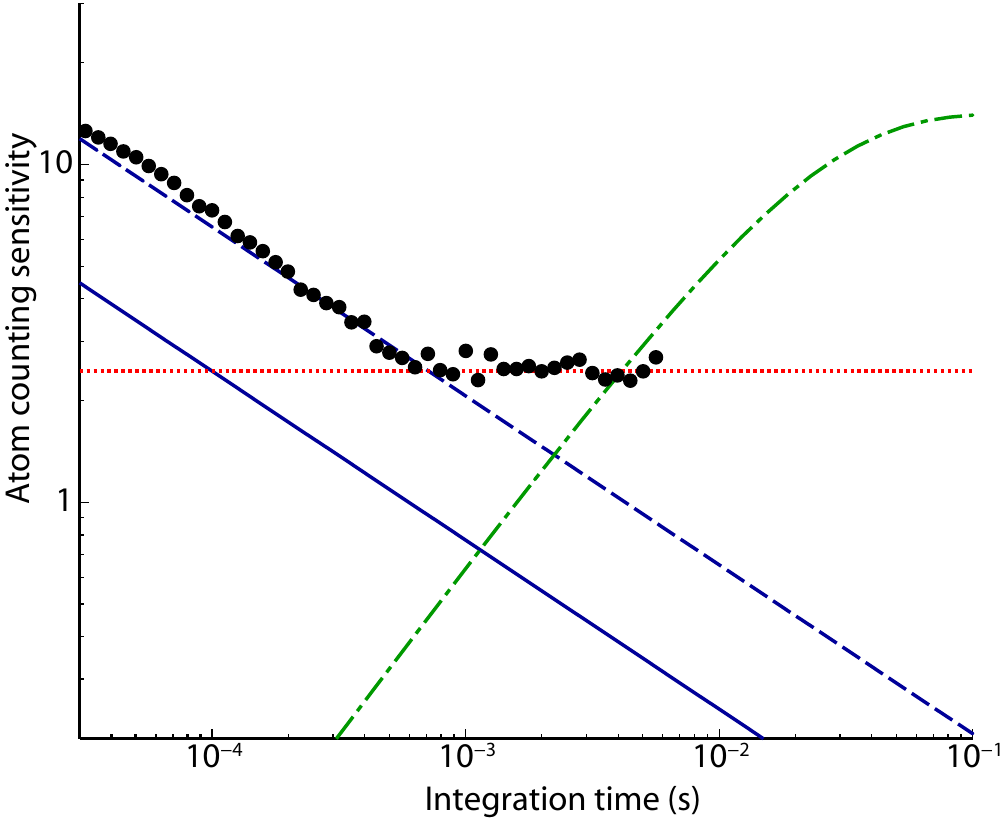}
	\caption{Atom-counting sensitivity at $\dca = -2$~GHz, with 200 atoms.  At low integration times the sensitivity is limited by shot noise (dashed blue line).  At near-unity detection efficiency, this limit would be $\sim 3$ times smaller (solid blue line).  At long integration times, the sensitivity is limited by technical fluctuations of the empty cavity resonance frequency (dotted red line).  At even longer integration times the counting sensitivity would be limited by fluctuations (green chain) in the average atom loss rate.  
The measured counting sensitivity is also shown (black dots).  Both scales are logarithmic.}
	\label{fig:sup:lowdcasens}
\end{figure}

Were this noise absent, the number-counting sensitivity would be limited by the atomic loss rate, which is due to radiation pressure fluctuations in the probe light.  The average loss is deterministic, and its effects can be corrected (cf.\ Supplementary Fig.~\ref{fig:sup:lowdcacount}).  However, fluctuations in the loss rate (consistent with fluctuations due from atomic shot noise) cause an uncertainty in this correction, and therefore limit the minimum counting uncertainty.



The atom counting uncertainty limits due to probe shot noise, atom loss, and technical fluctuations are shown in Supplementary Fig.~\ref{fig:sup:lowdcasens}.  Single-atom sensitivity would be obtained with near-unity detection efficiency and reduced empty-cavity resonance noise.


An example single-well counting measurement, with 60 atoms in a single lattice site, is shown in Supplementary Fig.~\ref{fig:sup:lowdcacount}.  The atom-counting sensitivity is $\pm 2.4$, using a 500~\tmu s integration time.

\begin{figure}[t]
	\centering
		\includegraphics{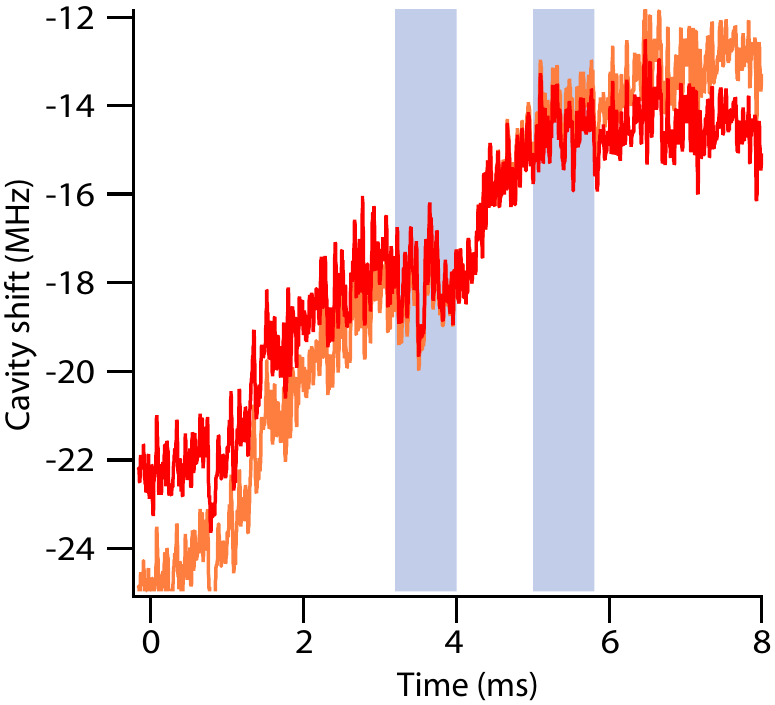}
	\caption{Single-site measurement at $\dca = -2$~GHz, with $\approx 200$ total atoms.  Measurement is taken by integrating the cavity shift, corrected for number loss (red line, normalized to the atom number at time 0), for 500~$\mu$s before and after flipping the spins in a single site (integration times shaded).  Here the spins are flipped from $|2,-2\rangle$ to $|2,2\rangle$ to minimize the initial loss.  The measurement yields $61.2\pm2.4$ atoms in the site at the time of the spin flip.  Also shown is the uncorrected shift (orange line).}
	\label{fig:sup:lowdcacount}
\end{figure}

\paragraph*{Transport.} To execute the transport experiment, we begin with atoms loaded into only a few lattice sites, with the atom distribution having a gaussian envelope of rms\ width $\sigma_0 = 380$~nm.  The atoms begin in a relatively deep optical lattice, with $U_0 = 84~E_r$.  The lattice depth is then lowered to $10.1~E_r$, and a weak field gradient ($\mu B'/h = 454$~Hz per lattice constant) is turned on to cancel gravity.  The atoms are allowed to tunnel for a time $t$, after which the gradient is raised to prevent resonant tunneling, and the lattice depth is raised to $410~E_r$.  At this point, the strong field gradient necessary for the MRI is applied, and an image is taken.

For resonantly tunneling non-interacting atoms we expect the position variance of the atom distribution to grow according to $\sigma_\mathrm{bal}^2 = \sigma_0^2 + \left ( a J t / \hbar \right )^2$, where $a$ is the lattice constant.  For the no-free-parameter fit of Fig.~\ref{fig:diffmri}, we determine $\sigma_0$ by taking an MRI without ever lowering the lattice.  We confirm this initial width by comparing the cavity shift as a function of the centre position of the atom distribution, effectively measuring the spatially varying position sensitivity of $\bar{g}$ (cf.\ Ref.~\citenum{purd10tunable}).  These two measurements of $\sigma_0$ agree to within the 15~nm statistical uncertainty of the latter method.
	\renewcommand{\refname}{Supplementary References}

\fi
}

\end{document}